\documentclass{kluwer}    

\usepackage{epsfig}
\newdisplay{guess}{Conjecture}

\begin{document}

\def\etal{{\sl et al.}}

\def\plotfiddle#1#2#3#4#5#6#7{\centering \leavevmode
\vbox to#2{\rule{0pt}{#2}}
\includegraphics{#1}}
%
%

\begin{article}
\begin{opening}         

\title{Galaxy Form and Spectral-Type: A Physical Framework for
Measuring Evolution}

\author{Matthew \surname{Bershady}}  
\runningauthor{Matthew Bershady}
\runningtitle{A Framework for Measuring Evolution}
\institute{University of Wisconsin-Madison}
\date{Sept 10, 1999}

\begin{abstract}

I outline a quantitative method for characterizing galaxies both by
photometric `form' and indices of spectral-type, applicable to both
nearby and distant galaxies. Such a characterization provides insight
on galaxy evolution because there are physical connections between
galaxies' stellar populations and their light distribution. ``Normal''
galaxies' form-parameters (surface-brightness, image concentration and
asymmetry) correlate well with spectral-index (color), which in turn
correlates only weakly with scale (size or luminosity). Deviations
from these normal relations also offer clues to the physical modes of
galaxy formation and evolution. As an example, I contrast a puzzling,
distant population of compact, but luminous, blue, star-forming
galaxies to nearby samples. These distant sources appear to be
associated with the bulk of the luminosity increase since $z<1$. They
have structural properties comparable to low-redshift populations, and
photometric properties within the norm for nearby, actively
star-forming galaxies. When combined, however, their photometric and
structural properties appear to be highly unusual.



\end{abstract}

\end{opening}           

\vskip -0.25in
\section{Classification and Evolution}


While great strides recently have been made in identifying galaxies
out to very large look-back times (e.g. Steidel \etal ~1999), when
galaxies form and how they evolve is still an issue of debate. The
primary physical processes determining a galaxy's appearance are the
aggregation of matter and the ensuing star-formation, stellar
evolution and chemical enrichment -- all within the dynamical
development of a gravitating system. It is possible and compelling to
model these processes and predict how galaxies appear {\it a priori}
(e.g. Contardo \etal ~1998). Our limited understanding of the
feed-back mechanisms associated with, and controlling star formation
make such simulations challenging. In a complementary approach, the
specific form and time-scales of these physical processes
(e.g. monolithic collapse vs. merging of bound systems, dissipation
vs. violent relaxation, and monotonic vs. stochastic star-formation
histories) may be differentiable within an observational
framework. Sandage (1986) outlined how the Hubble sequence today can
be interpreted in terms of different monolithic collapse and
star-formation histories. The interpretation, however, is not unique;
in hierarchical structure-formation scenarios different inferences are
drawn for how disks and spheroids form (e.g. van den Bausch 1998).
With a multiplicity of evolutionary paths to a single, current galaxy
type, an observational approach to differentiating between these paths
is desirable. {\it Since the ability to measure change relies on making a
comparison between galaxies at disparate distances, classification is
a cornerstone of galaxy evolution studies.}

To proceed observationally, a framework is needed which takes
advantage of current knowledge. For example, there is a {\it
fundamental} connection between stellar populations and local
galaxies' structure, known since Baade's study of on our own galaxy
and the bulge of M31. Simply interpreted, bulges are spatially
compact, nearly spherical, dynamically warm (V$_{\rm
rot}/\sigma\sim1$) systems composed of old, cool stars; disks are
dynamically cold (V$_{\rm rot}/\sigma\gg1$), more diffuse, and the
sites of recent, massive star-formation. This simple picture belies a
much more complex entanglement of stellar populations (e.g. King,
1971), and more varied dynamical and structural properties of disks
and bulges. Nonetheless, large galaxies today have discernibly
different spatial and dynamical distributions of stellar
populations. These differences are evolutionary clues which can be
exploited.

Here we consider the following physically motivated classification
scheme, to be applied to galaxies over a range of look-back times: The
amount of dissipation in bound, luminous matter is inferred directly
by measuring image concentration and surface-brightness. Time-scales
for star formation and matter aggregation are assessed independently
via characterization of stellar populations, gradients, changes in
galaxy scales, and asymmetry. Concentration, surface-brightness,
asymmetry, color, and luminosity (or size), then, compose at least
part of a critical subset of the classification tools necessary for
studying galaxy evolution. Together they are sensitive to the
temporal change in the spatial distribution of star-formation.

Such a classification scheme sounds remarkably similar to that
proposed over years ago by Morgan \& Mayall (1957). They reformulated
the concept of galaxy classification set forth by Hubble two decades
earlier, and made stellar populations the primary classification
parameter. This departure allowed Morgan (1958) to discover that the
spectral-type of galaxies' nuclear regions correlated strongly with
image concentration. (This is essentially the inverse of Hubble noting
that color correlated with morphological type, but here in more
physical terms of stellar densities and populations.)  Nonetheless,
image structure (morphology or ``form'') remained an essential
secondary parameter in Morgan's classification. {\it This indicates
the importance of both form and spectral-type as independent
classification axes.}

Form and spectral-indices\footnote{While Morgan intentionally defined
galaxy nuclear spectral-type within the paradigm of classical
stellar classification, this may not be suited ideally for a
general galaxy classification (e.g. admixtures of hot and cool stellar
components must be characterized; nuclear spectra may unobtainable or
poorly defined). For clarity, we henceforth refer to other measures of
galaxy spectral-type as spectral-indices.} are now being explored in
exquisite detail with modern data-sets; there has been much recent
progress in placing galaxy classification on a secure, quantitative
footing (e.g. these proceedings and references therein). Most of these
explorations, however, have stayed within the context of the Hubble
classification scheme, or consider either form or spectral-indices in
isolation. Conselice \etal~(2000) and Jangren \etal~(1999) recently
have demonstrated how modern measures of form and spectral-indices can
be used powerfully in concert. Whitmore (1984) long ago pointed the
way: His two dominant dimensions are ``form'' (disk-to-bulge ratio and
color) and ``scale'' (size and color). Whitmore's analysis, based on
local, luminous spirals, mixes what we term here as form and spectral
parameters. To develop a classification for both nearby and distant
samples we deliberately want to separate these parameters from each
other and from scale, and then determine how correlations between them
evolve.


Indeed, a crucial facet of classification is the ability to
incorporate galaxies of all scales (mass, size, and luminosity) in a
physically intelligible way. Can late-type galaxies, with neither
well-formed disks nor bulges but characteristically low luminosities,
be classified sensibly via form and spectral-indices alone?  
One distinguishing physical parameter here is the ratio of the current
to past-average star-formation rate. It should be possible to measure
this parameter not only spectroscopically (e.g. color), but via form,
through the amplitude of HII regions (flocculence, or high-frequency
asymmetry) relative to a smooth, underlying stellar population.  Hence
there is promise that form and color can be used to distinguish at
least between active and quiescent dwarf galaxies, and certainly
between the majority of dwarfs and giants observed
locally. Determining how such distinctions become distorted at higher
redshift gains us insight into how galaxies evolve.

\begin{figure}
\plotfiddle{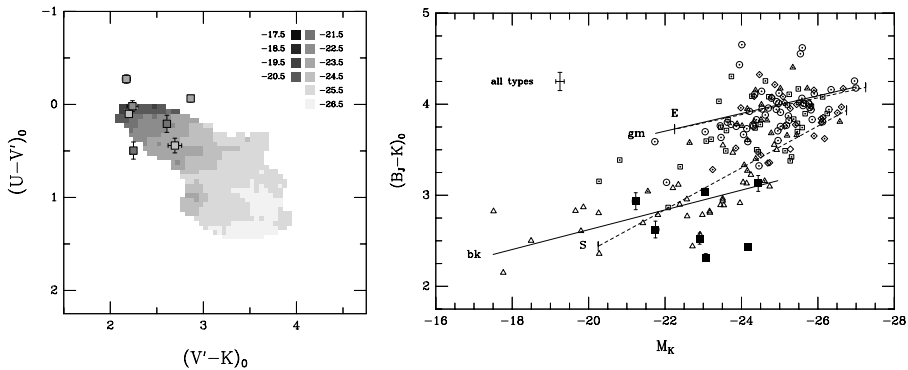}{1.85in}{0}{100}{100}{-307}{-500}
\vskip 0.05in
\caption{Left: $UV'K$ color-color diagram, showing the mean $K$-band
luminosity of a $B=20.5$ field galaxy survey as a function of
multi-color. Seven LBCGs are superimposed, shaded according to their
luminosity. Right: $B_{J}-K$ vs. M$_K$ color-magnitude diagram for the
same field sample, with the seven LBCGs again superimposed (dark
symbols). These figures, adapted from Bershady (1995; symbols, coded
by broad-band, optical--near-IR spectral-type are identified therein),
show that (a) there is a strong correlation in the mean between color
and luminosity for all galaxies, but a large range of luminosity for a
given color; (b) while LBCGs are among the most luminous galaxies for
their luminosity, they are not exceptional {\it as a class} when
compared to a general field population at low redshift.}
\vskip -0.1in
\end{figure}

In the balance of this paper I sketch some of the principal dimensions
of the classification motivated above, and demonstrate their utility
for understanding the nature of compact, luminous galaxies observed at
intermediate redshifts.

\vskip -0.15in
\section{A Modern Revision of Morgan's Classification Scheme}


There are numerous ways to define measures of form, spectral-index, and
scale. What is presented here is not unique, but the parameters have
been chosen to be robust to changes in signal-to-noise and image
resolution, and are cost-effective in terms of the required telescopic
observations relative to delivered information. Specifically, we use:

\begin{enumerate}

\item A spatially-integrated spectral-index, or stellar population
parameter: ideally measured via optical and near-infrared multi-colors
or spectra, but rest-frame $B-V$ can be used as a low-cost surrogate.

\item Form parameters: concentration (C), surface-brightness ($\mu$),
and asymmetry (A), as measured in a single, rest-frame band.

\item Scale parameters: size (r), and luminosity (L) -- these should
not be viewed as identical since there is a range of
surface-brightness at any given r or L. Ideally, a kinematic measure
of mass would be added as a third, independent scale parameter.

\end{enumerate}

\begin{figure}
\plotfiddle{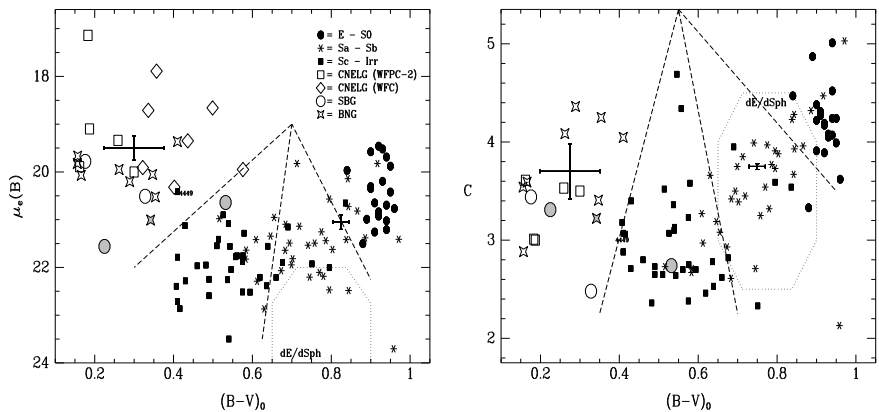}{1.9in}{0}{80}{100}{-240}{-515}
\vskip 0.1in
\caption{Left: $B$-band half-light surface-brightness
vs. $B-V$. Right: $B$-band concentration vs. $B-V$. Dashed lines are
heuristic classification boundaries, analogous to Abraham \etal
~(1996) for A vs C. Adapted from Jangren \etal ~(1999).}
\vskip -0.1in
\end{figure}

Figures 1-3 illustrate selected correlations between the above
parameters for two samples that we use to form a reference-set of
nearby galaxies: (i) the Frei \etal ~(1996) sample, as studied by
Conselice \etal ~(2000) and Jangren \etal ~(1999); and (ii) the
somewhat more distant ($z\sim0.13$) photometric sample from Bershady
(1995).  The latter contains a substantial number of low-redshift,
low-luminosity, actively star-forming systems, but has no measured
form-parameters. The Frei \etal ~sample, like most -- if not all --
bright galaxy samples, is unrepresentative of the number and variety
of low-luminosity and small systems. Consequently, the reference
distribution of form-parameters, based here only on the Frei \etal
~sample, should be viewed as preliminary. To amend this, we are in the
process of measuring form-parameters for other samples containing
nearby, star-forming dwarfs. Here, we schematically indicate the locus
of dwarf ellipticals/spheroidals in Figures 2 and 3 as compiled by
Jangren \etal ~(1999) and references therein. Finally, we have
included an intermediate redshift sample of what we will term
``luminous, blue compact galaxies,'' (LBCGs).

Focusing first on ``nearby'' samples, the following correlations are
apparent. Spectral-index and luminosity are well correlated (Figure
1), but there is a substantial range in luminosity for a given type
(color). The form parameters correlate strongly with spectral-index
(Figure 2), moderately well with each other (Figure 3), but very
weakly with scale ({\it N.B.} only large r and L probed here). To
complete this suite of figures, see e.g. Okamura \etal~(1984), or Kent
(1985) for $\mu$ vs. c; Conselice \etal~(2000), for A vs. $B-V$; and
Jangren \etal~(1999), for form vs. scale. While form and
spectral-index together distinguish well between normal Hubble types,
form-parameters alone are not as good -- particularly for separating
intermediate and late Hubble types. In other words, the dominant
difference between intermediate and late types, as defined by the
Hubble sequence, is color.

We note that while we are considering $\mu$ to be a form parameter, it
contains a hidden luminosity scale. This scale becomes evident if, for
example, the time-scales for size evolution is much larger than for
changes in the characteristic M/L a galaxy's stellar
population. Asymmetry and image concentration may also be affected by
evolution, but only by dynamical changes or spatially-dependent
variations in star-formation; $\mu$ is guaranteed to evolve even in a
dynamically relaxed and uniform stellar system. Hence it is important
to keep in mind that evolution may drive different changes in A, c and
$\mu$.

\begin{figure}
\plotfiddle{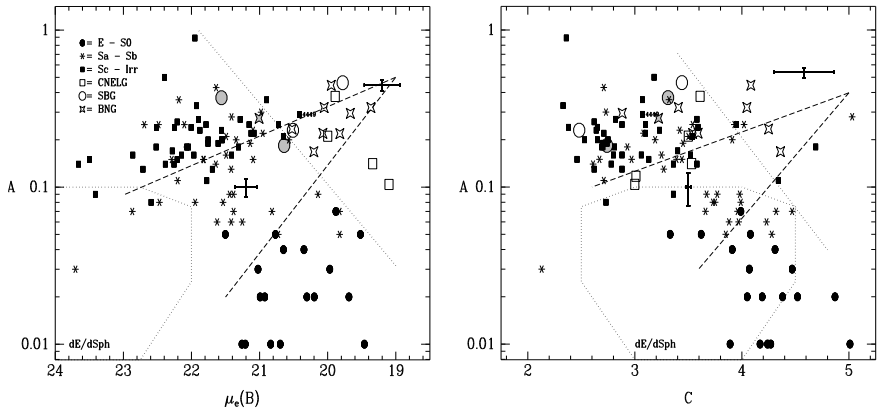}{2.15in}{0}{80}{100}{-240}{-500}
\caption{Left: $B$-band rotational asymmetry vs. half-light
surface-brightness. Right: $B$-band rotational asymmetry
vs. concentration.  Dashed lines are heuristic classification
boundaries, analogous to Abraham \etal ~(1996) for A vs C. Adapted
from Jangren \etal ~(1999).}
\vskip -0.1in
\end{figure}

\vskip -0.15in
\section{Luminous Blue Compact Galaxies}

A particularly intriguing distant population consists of small, but
luminous, blue, star-forming galaxies, found in a number of deep
redshift surveys. What we define as LBCGs (Jangren \etal ~1999)
contain the smallest galaxies for their observed luminosity (e.g. Koo
et al. 1994). Guzm\'an \etal ~(1997) and Lilly \etal ~(1998) have
argued that these sources appear to be the most rapidly evolving in
terms of space-density, and hence are likely to be associated with (if
not a key component of) the observed increase in star-formation
between $0<z<1$. How do they fit in to the above classification
scheme?

As seen in Figure 2, the LBCGs lie off the reference sequence when
viewed by form {\it and} spectral-index. However, their
optical--near-IR colors and luminosities are within the upper bounds
established for nearby field samples (Figure 1). In terms of
form-parameters alone (Figure 3), the LBCGs are extreme in A
vs. $\mu$, but not in terms of A vs. c.

Because our reference sample is incomplete, it is problematic to
assess how unusual the LBCGs are in form. Nonetheless, the most
luminous and compact LBCGs (a sub-class which we refer to as CNELGs)
do appear to be unusual even when compared to their slightly larger
counterparts (SBGs) of comparable rest-frame color and redshift
($z\sim0.4$). Figure 4 reveals color gradients (roughly between
rest-frame $U$ and $J$ bands) such that the most compact sources are
{\it bluer} in their centers, indicative of young, centrally
concentrated bursts embedded within older, more extended
populations. The slightly more extended sources, like nearby, normal
galaxies, show the characteristic reddening in their centers due to
the increasing relative dominance of the bulge. The most extreme
LBCGs, then, show the opposite correlation between form and stellar
type exhibited by galaxies along the Hubble sequence -- i.e., in
opposition to the underlying premise of our classification scheme!

How might these sources individually evolve? Are there local
counterparts? To start, the LBCGs' color gradients allow us to
conclude that the intermediate-$z$ bursts are not their first. More
difficult to ascertain is whether these sources (a) persist in, or
oscillate into and out of the luminous, blue phase via recurring
bursts, or (b) burst once or twice, and then fade into
obscurity. Certainly the stellar fossil record from the local group
(Grebel, 1998; Tolstoy, 1998) is consistent with multiple burst phases
for dwarfs. However, do the LBCGs have adequate gas to sustain
significant future bursts? {\it This critical question could be
answered by identifying comparable systems within redshift range of HI
observations} ($z\leq0.15$). If these sources have no subsequent
bursts and simply fade, their image structure will evolve to become
less concentrated, lower surface-brightness, and more symmetric. As in
previous photometric analysis (Guzm\'an \etal~1998), Jangren
\etal~(1999) find the faded colors {\it and} form of the LBCGs are
still consistent with some of today's spheroidal population
(e.g. NGC~205).

Finally, we would like to understand the origin of the
centrally-concentrated star-formation in these systems; it appears as
if they are forming ``outside-in.'' How was the gas which formed these
stars funneled into the central 1-2 kpc? There are tails and wisps in
these systems indicative of interactions or merging, but the bursts
are large (L$^*$ luminosities), may contain as much as 10\% of the
total stellar mass, and the total mass appears to be small
(M$\leq10^{10}$M$_\odot$) based on sizes and narrow line-widths. STIS
spectroscopy or high-angular resolution integral-field spectroscopy
would be invaluable to determine if the most extreme LBCGs are
rotationally supported, truly low mass, dynamically disturbed, or are
suffering from super-novae driven winds that may rid them of their ISM
and quench future star-formation.

{\bf Acknowledgements}: I would like to thank my collaborators whose
contributions made this work possible: A. Jangren, C. Conselice,
D. Koo, R. Guzm\'an, and C. Gronwall.  Funding for this work was
provided by STScI grants AR-07519, GO-07875 and NASA grant NAG5-6032.
\vskip -0.1in

\begin{figure}
\plotfiddle{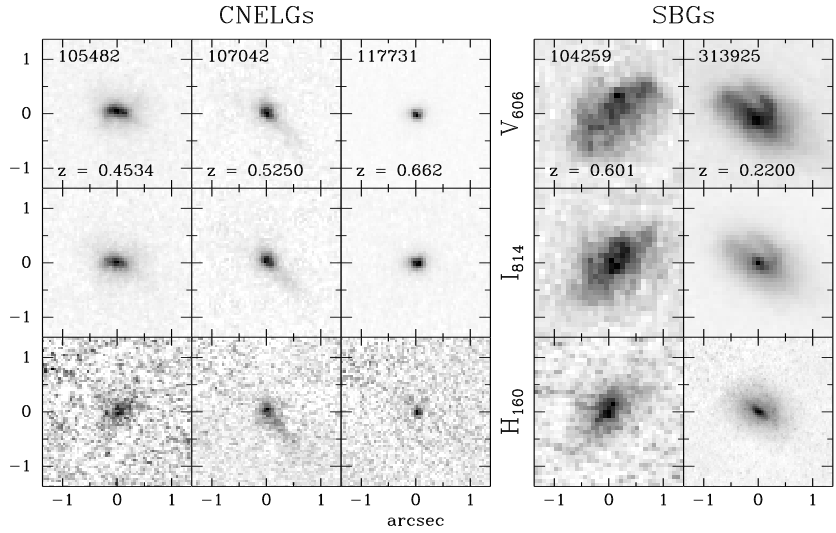}{2.75in}{-90}{45}{45}{-175}{260}
\caption{$VIH$ montage of three representative CNELGs and two SBGs
(see text) from our on-going HST WFPC-2 / NICMOS imaging programs of LBCGs. Note
the distinctly different systematic changes of apparent morpholgy with
band-pass between the CNELGs and SBGs: CNELGs are {\it more} compact
at shorter wavelengths.}
\end{figure}


\theendnotes

\end{article}

\begin{thebibliography}{}

\bibitem[\protect\citeauthoryear{Abraham \etal}{1996}]{abr96} 
Abraham, R.~G., \etal: 1996, MNRAS, 279, L47

\bibitem[\protect\citeauthoryear{Bershady}{1995}]{b95}
Bershady, M. A.: 1995, AJ, 109, 87

\bibitem[\protect\citeauthoryear{Conselice \etal}{2000}]{c00}
Conselice, C., Bershady, M. A., Jangren, A.: 2000, ApJ, in press (astro-ph/9907399)

\bibitem[\protect\citeauthoryear{Contardo \etal}{1998}]{cmu98}
Contardo, G., Steinmetz, M., Uta, F. A.: 1998, ApJ, 507, 497

\bibitem[\protect\citeauthoryear{Frei \etal}{1996}]{f96}
Frei, Z., Guhathakurta, P., Gunn, J.~E., Tyson, J.~A.: 1996, AJ, 111, 174 

\bibitem[\protect\citeauthoryear{Grebel}{1998}]{e98}
Grebel, E.: 1998, ASPCS, in press (astro-ph/9812443 )

\bibitem[\protect\citeauthoryear{Guzm\'an \etal}{1997}]{guz97} 
Guzm\'an, R., \etal: 1997, ApJ, 489, 559

\bibitem[\protect\citeauthoryear{Guzm\'an \etal}{1998}]{guz98} 
Guzm\'an, R., \etal: 1998, ApJ, 495, L13

\bibitem[\protect\citeauthoryear{Jangren \etal}{1999}]{j99}
Jangren, A., Bershady, M. A., Conselice, C.: 1999, AJ, submitted

\bibitem[\protect\citeauthoryear{Kent}{1985}]{k85}
Kent, S.~M.: 1985, ApJS, 59, 115

\bibitem[\protect\citeauthoryear{King}{1971}]{k71}
King, I. R.: 1971, PASP, 83, 377

\bibitem[\protect\citeauthoryear{Koo \etal}{1994}]{k94}
Koo, D.~C., \etal: 1994, ApJ, 427, L9

\bibitem[\protect\citeauthoryear{Lilly \etal}{1998}]{lil98} 
Lilly, S.~J., \etal: 1998, ApJ, 500, 75

\bibitem[\protect\citeauthoryear{Morgan and Mayall}{1957}]{mm57}
Morgan, W. W. \& Mayall, N. U.: 1957, PASP, 69, 291

\bibitem[\protect\citeauthoryear{Morgan}{1958}]{m58}
Morgan, W. W.
\newblock {\em PASP}, 70, 364, 1958

\bibitem[\protect\citeauthoryear{Okamura \etal}{1984}]{oka84}
Okamura, S., Kodaira, K., Watanabe, M.: 1984, ApJ, 280, 7 

\bibitem[\protect\citeauthoryear{Sandage}{1986}]{s86}
Sandage, A.: 1986, A\&A, 161, 89

\bibitem[\protect\citeauthoryear{Steidel \etal}{1999}]{s99} 
Steidel, C. C., \etal: 1999, ApJ, 519, 1


\bibitem[\protect\citeauthoryear{Tolstoy}{1999}]{t99}
Tolstoy, E.: 1999, in IAU Symposium 192, (astro-ph/9901245)

\bibitem[\protect\citeauthoryear{van den Bosch}{1998}]{v98}
van den Bosch, F. C.: 1998, ApJ, 507, 601

\bibitem[\protect\citeauthoryear{Whitmore}{1984}]{w84}
Whitmore, B. C.: 1984, ApJ, 278, 61


\end{thebibliography}
\end{document}